\documentclass[reprint,aps,prb,superscriptaddress]{revtex4-2}
\usepackage[dvipsnames]{xcolor}
\usepackage{dcolumn}
\usepackage{bm}
\usepackage[mathlines]{lineno}
\usepackage[utf8]{inputenc}
\usepackage{tikz}
\usepackage[T1]{fontenc}
\usepackage{booktabs, array, float, tabularx, lipsum, multirow, esvect}
\usepackage{paralist}
\usepackage{physics}

\usepackage{xr}
\usepackage{newtxtext,newtxmath}
\DeclareSymbolFont{orilargesymbols}{OMX}{cmex}{m}{n}
\DeclareMathSymbol{\orisum}{\mathop}{orilargesymbols}{"50}
\renewcommand\sum\orisum
\renewcommand\geq\geqslant
\renewcommand\leq\leqslant
\everymath{\displaystyle}
\usepackage{makecell}
\usepackage{graphicx}
\graphicspath{{../}}
\usepackage{siunitx}
\usepackage[version=4]{mhchem}
\usepackage[colorlinks,linkcolor=blue,anchorcolor=blue,citecolor=blue,urlcolor=blue]{hyperref}
\usepackage{orcidlink}
\makeatletter
\renewcommand{\figurename}{\textbf{FIGURE}}
\renewcommand{\tablename}{\textbf{TABLE}}
\renewcommand{\fnum@figure}{\figurename\ \textbf{\thefigure}}
\renewcommand{\fnum@table}{\tablename\ \textbf{\thetable}}
\makeatother
\usepackage{overpic}
\usepackage{setspace}
\begin{document}

\title{Nearly Complete Charge--Spin Conversion via Strain-Eliminated Fermi Pockets \\ in \texorpdfstring{$d$}{d}-Wave Altermagnets}
\newcommand{\WUSTa}{\affiliation{State Key Laboratory of Advanced Refractories, \href{https://ror.org/00e4hrk88}{Wuhan University of Science and Technology}, Wuhan 430081, People's Republic of China}}
\newcommand{\WUSTb}{\affiliation{School of Materials Science and Engineering, \href{https://ror.org/00e4hrk88}{Wuhan University of Science and Technology}, Wuhan 430081, People's Republic of China}}
\newcommand{\HBPU}{\affiliation{School of Electrical and Electronic Information Engineering, \href{https://ror.org/01z07eq06}{Hubei Polytechnic University}, Huangshi 435003, People's Republic of China}}
\newcommand{\WHU}{\affiliation{Key Laboratory of Artificial Micro- and Nano-structures of Ministry of Education,\\ School of Physics and Technology, \href{https://ror.org/033vjfk17}{Wuhan University}, Wuhan 430072, People's Republic of China}}
\newcommand{\YZU}{\affiliation{School of Physical Science and Technology, \href{https://ror.org/03tqb8s11}{Yangzhou University}, Yangzhou 225002, People's Republic of China}}

\author{Wancheng Zhang\orcidlink{0000-0003-1382-9806}}
\thanks{These authors contributed equally to this work}
\HBPU
\author{Fangqi Liu\orcidlink{0000-0001-9699-4713}}
\thanks{These authors contributed equally to this work}
\WHU
\author{Yong Liu\orcidlink{0000-0001-6095-6926}}
\WHU
\author{Jingguo Hu\orcidlink{0000-0003-3829-5177}}
\YZU
\author{Zhihong Lu\orcidlink{0000-0002-6636-0581}}
\email[Contact author:]{zludavid@live.com}
\WUSTb
\author{Rui Xiong\orcidlink{0000-0003-0468-6014}}
\email[Contact author:]{xiongrui@whu.edu.cn}
\WHU
\author{Zhenhua Zhang\orcidlink{0000-0003-2148-8239}}
\email[Contact author:]{zzhua@wust.edu.cn}
\WUSTb

\date{\today}

\begin{abstract}
Ideal $d$-wave altermagnets with nearly orthogonal flat Fermi surfaces enable complete spin-channel separation and 100\% theoretical charge--spin conversion efficiency (CSE). The metallic altermagnet \ce{KV2Se2O} exemplifies this, but realistic samples host residual elliptical Fermi pockets that enhance charge conductivity while suppressing spin conductivity, drastically reducing CSE. Here we show that in-plane equibiaxial tensile strain systematically eliminates these parasitic pockets, restoring the flat-band geometry. Our first-principles calculations reveal that CSE increases monotonically with strain, reaching a record $\sim$96\% at 4\% strain. An effective tight-binding model confirms that pocket suppression, governed by reduced next-nearest-neighbor hoppings, is the dominant mechanism. We further identify an unconventional out-of-plane spin current component with CSE $\sim$55\% at optimal orientations, enabling field-free perpendicular magnetization switching. Moreover, the same strain-driven removal of parasitic pockets yields giant TMR enhancement in \ce{KV2Se2O}-based magnetic tunnel junctions, from $10^{5}\%$ to $10^{9}\%$, and the giant TMR persists over a wide energy window near the Fermi level. These findings establish strain engineering as a clean, widely applicable strategy to maximize CSE and magnetoresistance in $d$-wave altermagnets, and provide predictive descriptors for screening high-efficiency spintronic materials.
\end{abstract}

\maketitle

\setlength{\parskip}{0pt}
\section{Introduction}\label{intro}
Altermagnetism represents a newly recognized class of collinear magnetic order
that combines the vanishing net magnetization of an antiferromagnet with a momentum-dependent spin
splitting reminiscent of a ferromagnet~\cite{smejkal_crystal_2020,mazin_editorial_2022,PhysRevX.12.031042,PhysRevX.12.040501,
vsmejkal2022anomalous,ma2021multifunctional,krempasky_altermagnetic_2024,bai_altermagnetism_2024}.
In contrast to conventional antiferromagnets, where opposite-spin sublattices are related by a
translation or inversion, altermagnets require a proper rotation or mirror operation to connect the
sublattices, giving rise to a nonrelativistic $d$-, $g$-, or $i$-wave spin splitting even in the
absence of spin-orbit coupling (SOC)~\cite{smejkal_crystal_2020,liu_spin-group_2022,guo_spin-split_2023}.
This unique symmetry enables the generation of time-reversal-odd ($\mathcal{T}$-odd) spin currents
with efficiencies that can far exceed those of conventional spin Hall materials, opening a
transformative avenue for next-generation spintronic memory and logic
devices~\cite{gonzalez-hernandez_efficient_2021,bai_efficient_2023,song_altermagnets_2025}.

A particularly compelling limit arises in $d$-wave altermagnets possessing two mutually orthogonal
flat Fermi surfaces with opposite spin polarizations~\cite{lai_d-wave_2025}. As captured by a minimal
two-band model, extreme spin-splitting anisotropy ($a/b\to 0$ or $\infty$) yields perfectly flat,
orthogonal Fermi sheets, corresponding to complete spin-channel separation in momentum space. Under
a longitudinal electric field along a $\langle100\rangle$ direction, a fully spin-polarized current
flows; under a field along a $\langle110\rangle$ direction, a pure transverse spin current emerges
with a magnitude equal to the total charge current. In this ideal configuration, the charge-to-spin
conversion efficiency (CSE) reaches its theoretical upper limit of $100\%$~\cite{lai_d-wave_2025}.

Recent experiments have identified the layered vanadium selenide oxide \ce{KV2Se2O} as a rare
metallic room-temperature $d$-wave altermagnet~\cite{jiang_metallic_2025}. The opposite-spin
sublattices are related by the $[C_2 \Vert C_{4z}]$ spin-space symmetry~\cite{smejkal_crystal_2020,
liu_spin-group_2022,mazin_editorial_2022}, which lifts Kramers
degeneracy without spin-orbit coupling and produces a giant momentum-dependent spin splitting
of $\sim 1.6$~eV at the $X$ and $Y$ points~\cite{jiang_metallic_2025,guo_spin-split_2023}.
First-principles calculations reveal that the pristine compound already exhibits a remarkable
CSE of $\sim 78\%$ at the charge neutrality point---nearly double that of the benchmark
altermagnet \ce{RuO2}---and can approach $98\%$ under modest electron doping, making it the
most efficient $\mathcal{T}$-odd spin-current generator reported to date~\cite{lai_d-wave_2025}.

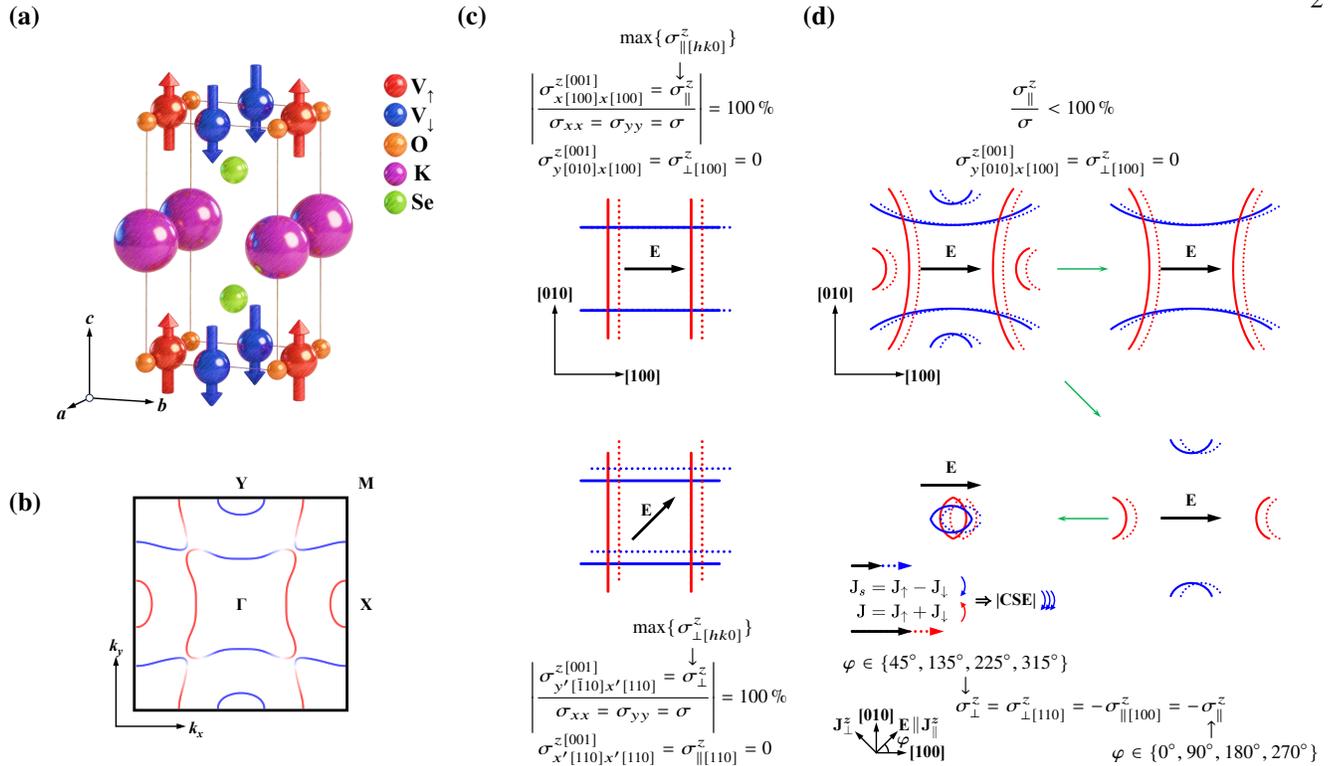
\begin{figure*}
  \centering
  \begin{overpic}[width=.99\textwidth]{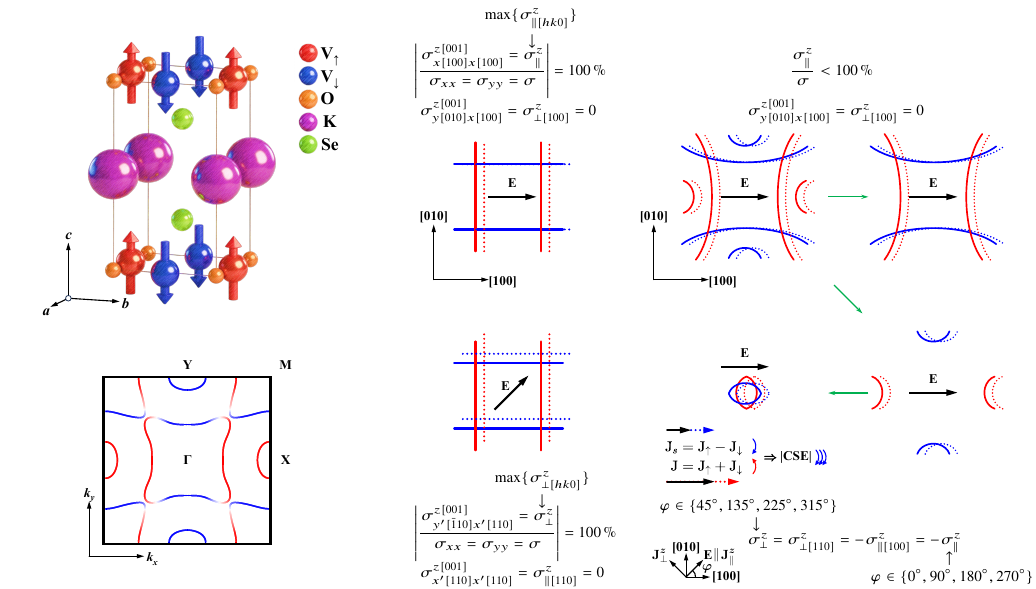}
    \put(0, 58){\normalsize \textbf{(a)}}
    \put(0, 20){\normalsize \textbf{(b)}}
    \put(35, 58){\normalsize \textbf{(c)}}
    \put(62, 58){\normalsize \textbf{(d)}}
  \end{overpic}
  \caption{Atomic structure, spin-resolved Fermi surface, and schematic of spin-current generation in \ce{KV2Se2O}.
    Red and blue denote opposite spin orientations (up and down, respectively).
    (a) Crystal structure and spin configuration of \ce{KV2Se2O}.
    (b) Spin-resolved Fermi surface in the $k_z = 0$ plane obtained from first-principles calculations.
    (c) Idealized spin-current generation.
    The spin-up and spin-down Fermi surfaces are perfectly orthogonal flat sheets.
    When an electric field $\mathbf{E}$ is applied along a $\langle 100 \rangle$ equivalent direction, a pure longitudinal spin-polarized current $\mathbf{J}_\parallel^z = \mathbf{J}_\uparrow$ is generated, quantified by the spin conductivity $\sigma_\parallel^z$.
    When $\mathbf{E}$ is applied along a $\langle 110 \rangle$ equivalent direction, a pure transverse spin current $\mathbf{J}_\perp^z = \mathbf{J}_\uparrow - \mathbf{J}_\downarrow$ emerges.
    Owing to the orthogonality of the flat bands, its magnitude satisfies $|\mathbf{J}_\perp^z| = |\mathbf{J}| = |\mathbf{J}_\uparrow + \mathbf{J}_\downarrow|$, and the corresponding spin conductivity is denoted $\sigma_\perp^z$.
    In this ideal limit, the maximum charge--spin conversion efficiency reaches $|\mathrm{CSE}|_{\mathrm{max}} = \sigma_\perp^z/\sigma = \sigma_\parallel^z/\sigma \equiv 100\%$, where $\sigma$ is the charge conductivity.
    (d) Realistic spin-current generation: residual Fermi pockets at $X$ and $Y$ points act as parasitic conduction channels that enhance charge conductivity while contributing oppositely to spin conductivity, drastically reducing the CSE relative to the ideal 100\% limit.
    \label{Fig1}}
\end{figure*}

Despite this record performance, pristine \ce{KV2Se2O} falls short of the ideal $100\%$ limit due
to the presence of residual elliptical Fermi pockets near the $X$ and $Y$ points. These pockets,
inherited from slight warping of the otherwise flat bands, act as parasitic conduction channels:
they enhance the charge conductivity while contributing oppositely to the spin conductivity, thereby
diluting the CSE. A natural strategy to eliminate these detrimental pockets
and recover the ideal flat-band geometry is strain engineering, which has already proven effective
in tuning altermagnetic properties in related materials such as \ce{RuO2}, \ce{MnTe}, \ce{CrSb} and
\ce{OsO2}~\cite{wang_strain-engineered_2026,PhysRevLett.134.086701,zhang_strain-induced_2025}.
While Lai \emph{et al.}~\cite{lai_d-wave_2025} achieved $\sim 98\%$ CSE through electron doping,
strain offers a more intrinsic route: it directly manipulates the lattice to restore the flat-band
geometry without introducing extrinsic carriers or altering the Fermi level. Chemical doping, in
contrast, inevitably introduces disorder and shifts the system away from charge neutrality. Strain
is a clean, continuously tunable parameter, readily accessible in epitaxial thin-film platforms and
compatible with conventional semiconductor processing~\cite{ZHAO2026100441,AFM-RuO2-S}.

In this work, we demonstrate through first-principles calculations [based on density functional
theory (DFT) plus Wannier interpolation; hereafter referred to as DFT calculations for brevity]
and effective-model analysis that in-plane equibiaxial tensile strain provides a highly effective
route to restore the ideal orthogonal flat Fermi surfaces in \ce{KV2Se2O}. The in-plane
$|\mathrm{CSE}|$ increases monotonically with strain, reaching a record value of $\sim 96\%$ at
$4\%$ strain, before slightly declining at larger deformations due to incipient Fermi-surface
bending. We further uncover an unconventional out-of-plane spin current component that achieves a
maximum $|\mathrm{CSE}_{\Uparrow}|$ of nearly $55\%$ at specific field orientations, offering a
promising pathway for field-free perpendicular magnetization switching. Beyond charge-to-spin
conversion, our magnetic-tunnel-junction (MTJ) calculations reveal that the strain-driven
elimination of parasitic Fermi pockets also yields a giant enhancement of tunneling
magnetoresistance (TMR), from $10^{5}\%$ to $10^{9}\%$, and this giant TMR persists over a wide
energy window near the Fermi level. Our findings establish strain engineering as a widely
applicable strategy to approach the ultimate CSE limit in $d$-wave altermagnets where parasitic
Fermi pockets limit the charge-to-spin conversion, and further demonstrate its efficacy in
simultaneously maximizing magnetoresistance, with direct implications for high-efficiency
spintronic devices.

\setlength {\parskip} {0pt}
\section{Results and Discussion}\label{rnd}

\begin{figure*}
  \centering
  \begin{overpic}[width=0.95\textwidth]{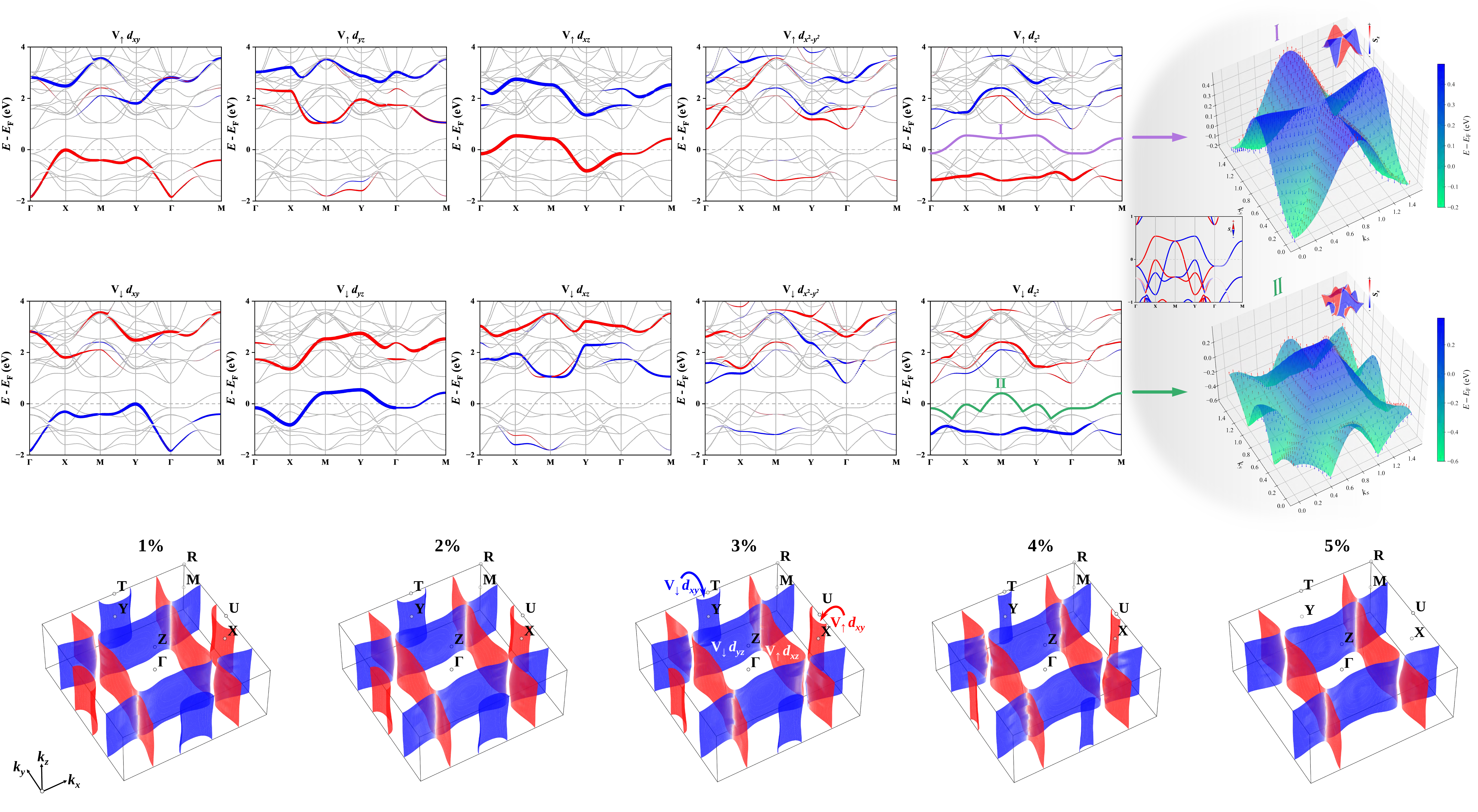}
    \put(0, 54){\normalsize \textbf{(a)}}
    \put(0, 18){\normalsize \textbf{(b)}}
    \put(78, 54){\normalsize \textbf{(c)}}
  \end{overpic}
  \caption{Band structure and Fermi surface of \ce{KV2Se2O} under equibiaxial tensile strain.
    (a) Calculated orbital- and spin-resolved band structures along high-symmetry lines at $4\%$ strain.
    Red and blue dots represent spin-up and spin-down bands, respectively.
    The size of the dots scales with the projection weight of the V $d_{xy}$, $d_{yz}$, $d_{xz}$, $d_{x^2-y^2}$, and $d_{z^2}$ orbitals.
    (b) Spin-resolved three-dimensional Fermi surfaces at $1\%$, $2\%$, $3\%$, $4\%$, and $5\%$ strain.
    (c) Spin texture of the two bands crossing the Fermi level [marked in purple and green in (a)] at $4\%$ strain. The arrows indicate the spin orientation,
    with red (blue) denoting positive (negative) out-of-plane spin component $S_z$.
    The surface color encodes the band energy. The inset shows the corresponding three-dimensional bands colored by $S_z$.
    Throughout the figure, red and blue represent spin-up and spin-down states, respectively.\label{Fig2}}
\end{figure*}

\subsection{Crystal Structure and Altermagnetic Configuration}

\ce{KV2Se2O} crystallizes in the tetragonal $P4/{\rm mmm}$ space group,
with spin-up and spin-down V atoms arranged within the same atomic plane.
The opposite-spin sublattices are related by the $[C_2 \Vert C_{4z}]$ symmetry operation,
as illustrated in Figure~\hyperref[Fig1]{\ref{Fig1}(a)}.
This unique magnetic configuration endows the material with pronounced crystalline and magnetic anisotropy,
enabling strong directional modulation of spin-dependent charge transport.
The spin-resolved Fermi surface in the $k_z=0$ plane, presented in Figure~\hyperref[Fig1]{\ref{Fig1}(b)},
directly reflects this anisotropic electronic structure: it exhibits the characteristic $d$-wave altermagnetic
anisotropy~\cite{smejkal_crystal_2020,mazin_editorial_2022}, wherein spin-up and spin-down states are
predominantly orthogonal yet retain small residual pockets near the $X$ and $Y$ points that break the perfect flat-band limit.

The idealized spin-current generation mechanism is illustrated in
Figure~\hyperref[Fig1]{\ref{Fig1}(c)}: in the perfect $d$-wave altermagnetic limit,
the spin-up and spin-down Fermi surfaces form mutually orthogonal flat sheets.
An electric field along $\langle 100\rangle$ drives a purely longitudinal
spin-polarized current, while a field along $\langle 110\rangle$ generates a pure
transverse spin current whose magnitude equals the total charge current, yielding
a theoretical $|\mathrm{CSE}|$ of $100\%$. In reality, as depicted in
Figure~\hyperref[Fig1]{\ref{Fig1}(d)}, residual elliptical pockets near $X$ and $Y$
act as parasitic conduction channels that enhance the charge conductivity while
contributing oppositely to the spin conductivity, thereby reducing the CSE far
below the ideal limit.

\subsection{Strain-Modulated Electronic Structure}

The energetic feasibility of the strain-engineering strategy is confirmed by DFT
calculations with SOC: the strain energy follows a quadratic dependence
$\Delta E \propto \varepsilon^2$ characteristic of the elastic regime, with a modest
cost of only $\sim\SI{245}{\milli\electronvolt}$ per conventional cell at $4\%$ strain,
and no structural instabilities are observed across the full 0\%--5\% range
(Supplemental Material~\cite{supplement}, Figure~\ref{fig:bands}).

Figure~\hyperref[Fig2]{\ref{Fig2}(a)} displays the orbital- and spin-resolved band
structure of \ce{KV2Se2O} at $4\%$ equibiaxial tensile strain (see band structures at all strain in Supplemental Material~\cite{supplement}).
The overall band dispersion and the pronounced spin splitting near the $X$ and $Y$ points are in
excellent agreement with the results reported by Jiang
\emph{et al.}~\cite{jiang_metallic_2025}. From Figures~\hyperref[Fig2]{\ref{Fig2}(a)}
and \hyperref[Fig2]{\ref{Fig2}(b)} it is evident that the mutually orthogonal flat
bands---the primary source of the large CSE---are predominantly derived from the
spin-up V $d_{xz}$ and spin-down V $d_{yz}$ orbitals, oriented nearly along the
$k_y$ and $k_x$ directions, respectively. The elliptical pockets emerging near $X$
and $Y$, on the other hand, originate from the spin-up and spin-down V $d_{xy}$
orbitals. As illustrated schematically in Figure~\hyperref[Fig1]{\ref{Fig1}(d)},
these pockets act as parasitic conduction channels that suppress the spin
conductivity and therefore limit the overall CSE.

Figure~\hyperref[Fig2]{\ref{Fig2}(b)} presents the spin-resolved three-dimensional Fermi
surfaces at strain levels of 1\%, 2\%, 3\%, 4\%, and 5\%. The progressive elimination
of the elliptical pockets is clearly visible: at 1\% strain the pockets are still
prominent; by 3\% they have significantly diminished; and at 4\% strain they are nearly
completely suppressed, leaving only the orthogonal flat sheets that characterize the
ideal $d$-wave altermagnetic limit. Beyond 4\%, slight bending of the previously flat
sheets begins to emerge, which correlates with the slight decline in CSE observed at
5\% strain (Table~\ref{tab:strain_transport}).

The distinctive $d$-wave flat Fermi surface of \ce{KV2Se2O} and its
strain-enhanced charge-to-spin conversion can be captured by an effective
two-orbital tight-binding model (four bands in total). In the absence of
spin-orbit coupling, the Hamiltonian decouples into spin-up and spin-down
sectors related by the altermagnetic $[C_2 \Vert C_{4z}]$ symmetry.
For each spin projection, we adopt a $2\times2$ low-energy basis that
phenomenologically describes the flat sheet (derived from the $d_{xz}$/$d_{yz}$
orbitals) and the elliptical pocket (originating from the $d_{xy}$ orbital).
Fitting to the DFT band structure yields the hopping parameters
listed in the Supplemental Material~\cite{supplement}. The model naturally
reproduces the orthogonal flat Fermi segments along $k_x$ and $k_y$, as well
as the isolated pockets at $X$ and $Y$, providing a transparent framework for
analyzing the strain-driven evolution of the CSE.

The spin texture of the two bands crossing the Fermi level at $4\%$ strain is shown
in Figure~\hyperref[Fig2]{\ref{Fig2}(c)}. The arrows, colored by the out-of-plane
spin component $S_z$, point predominantly along the $z$ direction, indicating that
the spin polarization is almost entirely out-of-plane. A clear $d$-wave pattern
emerges in the sign distribution of $S_z$: positive and negative values are
separated by the nodal lines $k_x=\pm k_y$, across which the spin polarization
abruptly reverses~\cite{d-wave1,d-wave2}. Along a fixed band index, this reversal is particularly sharp
in the vicinity of the elliptical pockets at $X$ and $Y$, where the DFT eigenvalues
switch between the spin-up and spin-down branches of the $d_{x^2-y^2}$-wave spin
splitting. Within the effective tight-binding model, the spin splitting is
dominated by the $(\cos k_x - \cos k_y)$ term, which vanishes along the diagonals
and attains opposite signs on either side of the pockets (a detailed symmetry
analysis is provided in the Supplemental Material~\cite{supplement}). The sign
change along the band index therefore maps directly onto the nodal structure of
the $d$-wave order, confirming that the pocket regions act as momentum-space
domain boundaries where the spin character of the highest valence bands is
exchanged. This observation further corroborates the altermagnetic nature of
\ce{KV2Se2O} and highlights the intimate connection between Fermi surface
topology and spin-dependent transport.

\subsection{Charge-to-Spin Conversion Efficiency under Strain}

\begin{figure*}
  \centering
  \begin{overpic}[width=0.95\textwidth]{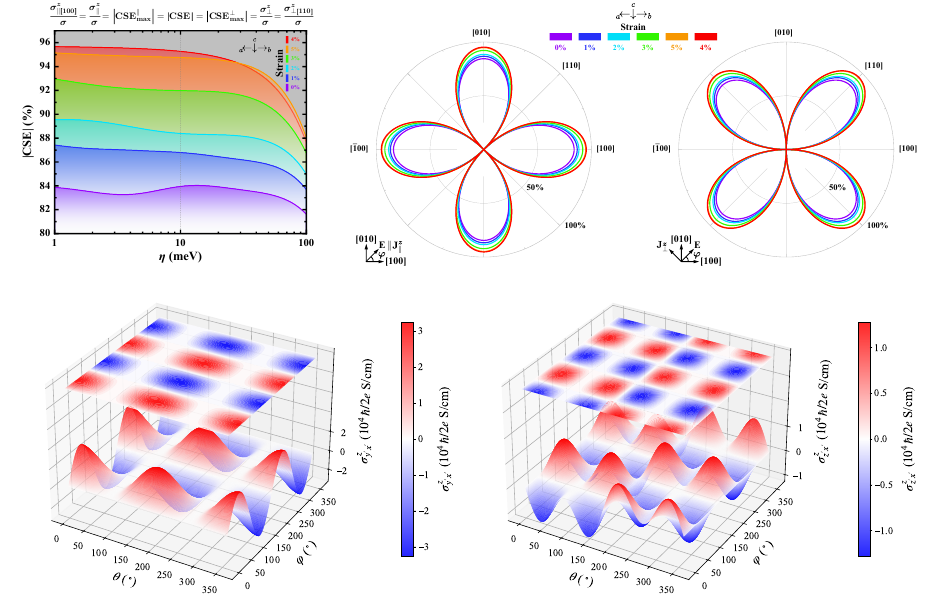}
    \put(2, 62){\normalsize \textbf{(a)}}
    \put(40, 62){\normalsize \textbf{(b)}}
    \put(2, 30){\normalsize \textbf{(c)}}
    \put(50, 30){\normalsize \textbf{(d)}}
  \end{overpic}
  \caption{Strain and angular dependence of the charge-to-spin conversion efficiency (CSE) and
      spin conductivities in \ce{KV2Se2O} evaluated with a broadening parameter $\eta = 10$~meV.
      (a) Strain-dependent CSE as a function of $\eta$.
      (b) Polar plots of the longitudinal (left) and transverse (right) CSE under an
      in-plane rotating electric field at various strain levels. The fourfold symmetry
      and the relation $\mathrm{CSE}_{\perp}(\varphi=\SI{45}{\degree}) = -\mathrm{CSE}_{\parallel}(\varphi=\SI{0}{\degree})$
      are direct consequences of the $d$-wave altermagnetic order.
      (c) Full-angle dependence of the transverse spin conductivity
      $\sigma_{y'x'}^z$ at $0\%$ strain, shown as a 3D surface combined with a
      top-view contour projection.
      (d) Corresponding out-of-plane spin conductivity $\sigma_{z'x'}^z$ displayed
      in the same 3D-plus-contour representation.
      \label{Fig3}}
\end{figure*}
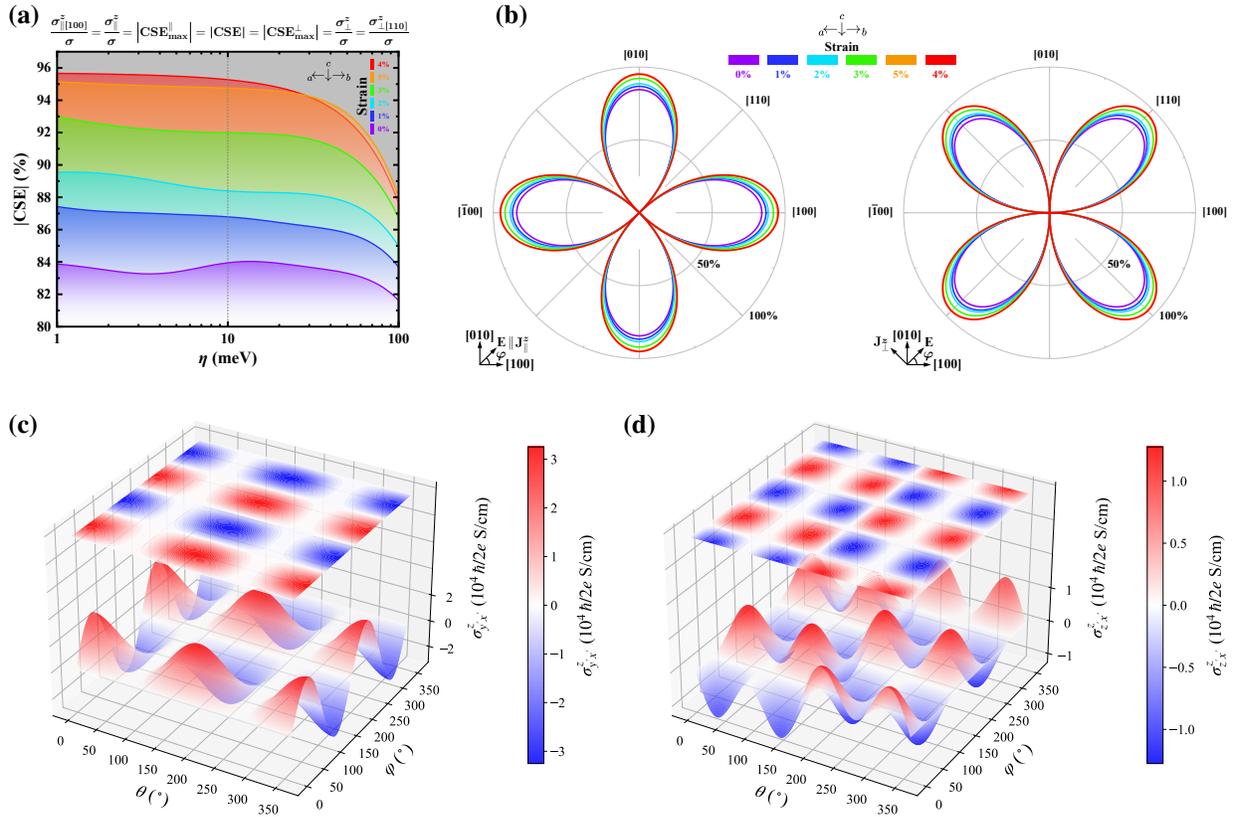

\begin{table*}[t]
  \caption{Strain dependence of the charge conductivity $\sigma$, the in-plane spin conductivity $|\sigma_{\vdash}^s| = |\sigma_{\parallel}^z| = |\sigma_{\perp}^z|$,
    the out-of-plane spin conductivity $|\sigma_{\Uparrow}^{\hat{\mathbf{n}}}|$, and the corresponding charge-to-spin conversion
    efficiencies $|\mathrm{CSE}|$ (in-plane) and $|\mathrm{CSE}_{\Uparrow}|$ (out-of-plane) for \ce{KV2Se2O}.}
    \label{tab:strain_transport}
    \begin{ruledtabular}
      \begin{tabular}{ccccc}
        Strain (\%) & $\sigma$ (S/cm) & $|\sigma_{\vdash}^s|$ [$(\hbar/2e)$~S/cm] & $|\sigma_{\Uparrow}^{\hat{\mathbf{n}}}|$ [$(\hbar/2e)$~S/cm] & |CSE| (|CSE$_{\Uparrow}$|) (\%) \\ \hline \\ [-2ex]
             0      & $3.78 \times 10^4$ & $3.17 \times 10^4$ & $1.28 \times 10^4$ &  83.95 (48.56)  \\
             1      & $3.56 \times 10^4$ & $3.09 \times 10^4$ & $1.20 \times 10^4$ &  86.79 (49.67)  \\
             2      & $3.06 \times 10^4$ & $2.71 \times 10^4$ & $1.10 \times 10^4$ &  88.39 (50.51)  \\
             3      & $2.80 \times 10^4$ & $2.58 \times 10^4$ & $1.02 \times 10^4$ &  91.98 (52.64)  \\
             4      & $2.71 \times 10^4$ & $2.58 \times 10^4$ & $1.01 \times 10^4$ &  95.28 (54.34)  \\
             5      & $2.59 \times 10^4$ & $2.46 \times 10^4$ & $0.96 \times 10^4$ &  94.75 (54.10)  \\
      \end{tabular}
    \end{ruledtabular}
\end{table*}

The altermagnetic symmetry not only dictates the Fermi surface topology but also
governs the angular dependence of the spin currents. To quantify this, we compute
the longitudinal and transverse CSEs as the electric field rotates within the $xy$
plane. As shown in Figure~\hyperref[Fig3]{\ref{Fig3}(b)}, the longitudinal CSE
exhibits a fourfold symmetric pattern with maxima along $\langle100\rangle$
directions, while the transverse CSE peaks along $\langle110\rangle$ and vanishes
along $\langle100\rangle$, consistent with the $d$-wave character of the spin
splitting. The two satisfy the exact relation
$|\mathrm{CSE}_{\perp}|(\varphi = \SI{45}{\degree}) = |\mathrm{CSE}_{\parallel}|(\varphi = \SI{0}{\degree})$,
inherited from the $[C_2 \Vert C_{4z}]$ symmetry.

Figure~\hyperref[Fig3]{\ref{Fig3}(a)} shows the computed CSE as a function of the
broadening parameter $\eta$ at different strain levels. The CSE exhibits a clear
monotonic increase with strain across the entire range of $\eta$, confirming the
robustness of the strain-enhancement effect. Quantitatively, at the optimal $4\%$
strain, the CSE varies by less than 0.9 percentage points within the physically
relevant range $\eta = 1$--$20$~meV, and the $\eta$ threshold at which the CSE
deviates by more than 1 percentage point from its $\eta=10$~meV reference value exceeds $50$~meV.
A systematic $\eta$-sensitivity analysis across all strain levels is provided in
the Supplemental Material~\cite{supplement}.

Table~\ref{tab:strain_transport} summarizes the strain evolution of the charge
conductivity, the in-plane and out-of-plane spin conductivities, and the corresponding
CSEs. With increasing tensile strain, the charge conductivity $\sigma$ decreases
monotonically from $3.78\times 10^4$ to $2.59\times 10^4$~S/cm, reflecting the
progressive removal of the parasitic pocket spectral weight. The in-plane spin
conductivity $|\sigma_{\vdash}^s|$ decreases from $3.17\times 10^4$ to
$2.46\times 10^4$~$(\hbar/2e)$~S/cm, while the in-plane $|\mathrm{CSE}|$ rises
from $\sim 84\%$ to a peak value of $95.28\%$ at $4\%$ strain.
The slight difference from the $\sim 78\%$ reported by Lai \emph{et al.}~\cite{lai_d-wave_2025} 
for pristine \ce{KV2Se2O} originates from the use of Hubbard $U$ corrections~\cite{jiang_metallic_2025,Hubbard_U1,Hubbard_U2} 
and different pseudopotentials.
The out-of-plane spin conductivity
$|\sigma_{\Uparrow}^{\hat{\mathbf{n}}}|$ (discussed in detail in Sec.~\ref{oop}) exhibits a similar downward trend, dropping
from $1.28\times 10^4$ to $0.96\times 10^4$~$(\hbar/2e)$~S/cm, yet its associated
conversion efficiency $|\mathrm{CSE}_{\Uparrow}|$ increases from $48.56\%$ at
$0\%$ strain to a peak of $54.34\%$ at $4\%$ strain, before a slight decline at
$5\%$ strain.

Two notable features in Table~\ref{tab:strain_transport} warrant closer examination:
the plateau in $|\sigma_{\vdash}^s|$ between 3\% and 4\% strain, and the slight
decline of the in-plane CSE from 4\% to 5\%. By 3\% strain, the elliptical pockets
have been largely eliminated; the residual pocket contribution to both $\sigma$ and
$\sigma^s$ is already minimal. Between 3\% and 4\%, the spin conductivity therefore
remains nearly constant at $2.58\times10^4$~$(\hbar/2e)$~S/cm as it approaches the
intrinsic limit set by the orthogonal flat sheets alone. The charge conductivity
$\sigma$, however, continues to decrease because the vanishing pocket DOS reduces
the total spectral weight at the Fermi level. Consequently, the CSE continues to
rise ($91.98\%\to 95.28\%$) despite the plateau in $|\sigma_{\vdash}^s|$. At 5\%
strain, the previously flat Fermi sheets begin to develop a slight curvature [see
Figure~\hyperref[Fig2]{\ref{Fig2}(b)}], which has two competing effects: it increases
the Fermi velocity on the formerly flat sheets, modestly enhancing the charge
conductivity, while simultaneously reducing the perfect nesting that maximizes the
$\mathcal{T}$-odd spin current. The net result is a small drop in CSE from
$95.28\%$ to $94.75\%$, marking 4\% as the optimal strain where pocket elimination
is maximal before flat-band degradation sets in. This non-monotonic behavior reveals
a fundamental trade-off: sufficient tensile strain is required to suppress the
parasitic pockets, but excessive strain eventually distorts the flat bands
themselves---a generic feature of the pocket-elimination mechanism in $d$-wave
altermagnets.

\begin{table}[htbp]
  \caption{Strain-dependent optimal orientations and the corresponding
    $\mathrm{CSE}_{\mathrm{or}}^{\hat{\mathbf{n}}}$
    in \ce{KV2Se2O}. For the $(011)$ orientation, the new $x'$ axis (electric-field
    direction) is $[01\bar{1}]$ and the new $y'$ axis is $[\bar{1}00]$; for $(111)$,
    $x' = [0\bar{1}1]$ and $y' = [2\bar{1}1]$. The angles $\theta$ and $\varphi$ define
    the direction of $\mathbf{E}$ in the original crystal frame.}
  \label{tab:orientation}
  \begin{ruledtabular}
    \begin{tabular}{ccccc}
      \multicolumn{1}{c}{Strain (\%)} & Orientation & $\theta$ ($^\circ$) & $\varphi$ ($^\circ$) & $\mathrm{CSE}_{\mathrm{or}}^{\hat{\mathbf{n}}}$ (\%) \\
      \hline \\ [-2ex]
      \multirow{2}{*}{0}              & $(011)$     & $151.98$              & $90.00 $               & $39.74$ \\
                                      & $(111)$     & $ 28.02$              & $270.00$               & $39.74$ \\[2pt]
      \multirow{2}{*}{1}              & $(011)$     & $150.75$              & $90.00 $               & $41.88$ \\
                                      & $(111)$     & $ 29.25$              & $270.00$               & $41.88$ \\[2pt]
      \multirow{2}{*}{2}              & $(011)$     & $150.27$              & $90.00 $               & $44.02$ \\
                                      & $(111)$     & $ 29.73$              & $270.00$               & $44.02$ \\[2pt]
      \multirow{2}{*}{3}              & $(011)$     & $149.44$              & $90.00 $               & $47.14$ \\
                                      & $(111)$     & $ 30.56$              & $270.00$               & $47.14$ \\[2pt]
      \multirow{2}{*}{4}              & $(011)$     & $148.64$              & $90.00 $               & $48.83$ \\
                                      & $(111)$     & $ 31.36$              & $270.00$               & $48.83$ \\[2pt]
      \multirow{2}{*}{5}              & $(011)$     & $147.86$              & $90.00 $               & $50.01$ \\
                                      & $(111)$     & $ 32.14$              & $270.00$               & $50.01$
    \end{tabular}
  \end{ruledtabular}
\end{table}

\subsection{Out-of-Plane Spin Current and Optimal Orientations}\label{oop}

When the electric field is allowed to tilt out of the $ab$ plane, an unconventional
out-of-plane spin current emerges. Figures~\hyperref[Fig3]{\ref{Fig3}(c)} and
\hyperref[Fig3]{\ref{Fig3}(d)} display the full angular dependence of the
transverse in-plane and out-of-plane spin conductivities over the entire solid angle
$(\theta,\varphi)$. The in-plane transverse component dominates near the equatorial
plane ($\theta = \SI{90}{\degree}$) and exhibits the same fourfold oscillation as in
the strictly two-dimensional case. For convenience, we denote the maximum in-plane
(longitudinal/transverse) spin conductivity by $\sigma_{\vdash}^s$ and the
corresponding maximum charge-to-spin conversion efficiency by CSE.
In contrast, the out-of-plane spin conductivity $\sigma_{z'x'}^z$
vanishes identically within cones centered on the $c$ axis ($\theta \approx \SI{90}{\degree}$
and $\SI{270}{\degree}$, half-angle $\sim \SI{3}{\degree}$), and reaches maximal magnitude (hereafter denoted
$\sigma_{\Uparrow}^{\hat{\mathbf{n}}}$, with $\hat{\mathbf{n}}$ being the N\'eel vector direction) when
the field is tilted by $\sim \SI{35}{\degree}$ toward $\langle 100\rangle$ or $\langle 010\rangle$,
corresponding closely to the $\langle 101\rangle$ and $\langle 011\rangle$ crystal directions.
The symmetry-enforced relation $\sigma_{\Uparrow}^{\hat{\mathbf{n}}}(\varphi+\SI{90}{\degree})
= -\sigma_{\Uparrow}^{\hat{\mathbf{n}}}(\varphi)$ is strictly observed for the signed conductivity,
confirming the $[C_2\parallel C_{4z}]$ altermagnetic order.
The corresponding global maximum out-of-plane CSE, denoted
$\mathrm{CSE}_{\Uparrow}^{\hat{\mathbf{n}}}$ and listed in
Table~\ref{tab:strain_transport}, reaches $54.34\%$ at $4\%$ strain---nearly $55\%$---following
the same optimal strain as the in-plane component.

Table~\ref{tab:orientation} lists the optimal crystallographic orientations at each
strain level and reports an orientation-specific metric
$\mathrm{CSE}_{\mathrm{or}}^{\hat{\mathbf{n}}}$ evaluated at the experimentally
accessible $(011)$ and $(111)$ directions. While the global maximum
$\mathrm{CSE}_{\Uparrow}^{\hat{\mathbf{n}}}$ peaks at $4\%$ strain
(Table~\ref{tab:strain_transport}), $\mathrm{CSE}_{\mathrm{or}}^{\hat{\mathbf{n}}}$
continues to increase up to $5\%$, reaching $50.01\%$, because the
orientation-specific optimum shifts slightly with strain. A complete list of extrema,
their crystallographic assignments, and a detailed discussion of the rotation
formalism are provided in the Supplemental Material~\cite{supplement}.

\subsection{Effective Tight-Binding Model Analysis}

\begin{figure*}[htbp]
\centering
\begin{overpic}[width=0.95\textwidth]{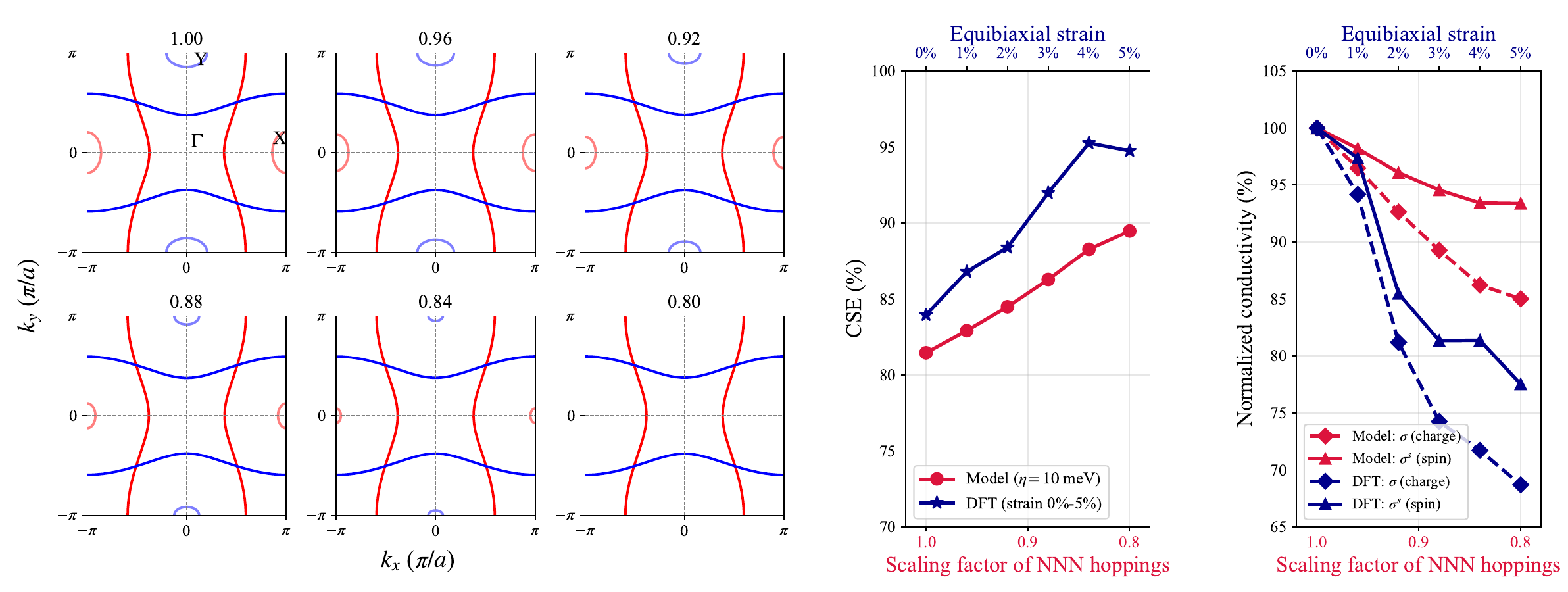}
    \put(0, 35){\normalsize \textbf{(a)}}
    \put(52, 35){\normalsize \textbf{(b)}}
    \put(77, 35){\normalsize \textbf{(c)}}
\end{overpic}
\caption{Strain-induced evolution of the Fermi surface and transport properties in the
effective tight-binding model for \ce{KV2Se2O}.
(a) Spin-resolved Fermi surfaces in the $k_z=0$ plane as the NNN
hopping parameters are uniformly scaled from $1.00$ (pristine) to $0.80$ in steps of
$0.04$. Red and blue contours correspond to spin-up and spin-down bands, respectively.
(b) CSE evaluated from the Kubo formula with broadening $\eta = 10$~meV
and from DFT results ($\eta = 10$~meV). The CSE increases monotonically
from $\sim 82\%$ to $\sim 90\%$ as the pockets are suppressed.
(c) Normalized charge conductivity $\sigma$ and spin conductivity $\sigma^s$.
Solid and dashed lines represent model and DFT results, respectively.
}
\label{fig:model_FS_CSE_evolution}
\end{figure*}

To establish a direct link between the microscopic hopping parameters and the
strain-enhanced CSE, we employ the effective tight-binding model fitted to the
pristine DFT bands (see Supplemental Material~\cite{supplement} for the full
Hamiltonian). Figure~\hyperref[fig:model_FS_CSE_evolution]{\ref{fig:model_FS_CSE_evolution}(a)}
displays the evolution of the spin-resolved Fermi surface when the
next-nearest-neighbor (NNN) hopping terms responsible for the elliptical
pockets---i.e., those proportional to $\cos k_x\cos k_y$, $\cos 2k_x$, and
$\cos 2k_y$ in the Hamiltonian---are uniformly scaled by a factor ranging from
$1.00$ (pristine) to $0.80$ in steps of $0.04$. This global suppression mimics
the reduction of longer-range hoppings under in-plane equibiaxial tensile strain.
As the scaling factor decreases, the elliptical pockets at $X$ and $Y$ gradually
shrink and ultimately vanish, leaving only the orthogonal flat sheets
characteristic of the ideal $d$-wave altermagnetic limit.

Figure~\hyperref[fig:model_FS_CSE_evolution]{\ref{fig:model_FS_CSE_evolution}(b)}
shows the corresponding CSE computed from the Kubo formula with a constant
broadening $\eta = 10$~meV, alongside the CSE values obtained from explicit DFT
strain calculations. The model CSE rises monotonically from $\sim 82\%$ to
$\sim 90\%$, closely following the DFT trend and confirming that pocket
elimination is the dominant mechanism for the strain-induced enhancement.

The origin of this enhancement is revealed in
Figure~\hyperref[fig:model_FS_CSE_evolution]{\ref{fig:model_FS_CSE_evolution}(c)},
where we plot the normalized charge conductivity $\sigma$ and spin conductivity
$\sigma^s$. Both conductivities decrease as the pockets are suppressed, but
$\sigma$ drops at a markedly faster rate. This is a direct consequence of the
pockets acting as conductive channels that contribute constructively to the
charge current yet destructively to the spin current. Their removal therefore
increases the CSE, driving the system toward the ideal flat-band limit. The same
trend is consistently observed in the DFT transport data listed in
Table~\ref{tab:strain_transport}: as the tensile strain increases from $0\%$ to
$5\%$, the charge conductivity $\sigma$ declines monotonically while the spin
conductivity $\sigma^s$ plateaus between $3\%$ and $4\%$ strain (after the pockets
are largely eliminated), enabling the CSE to climb from $\sim 84\%$ to $\sim 96\%$ at
$4\%$ strain. This quantitative agreement
between the model and DFT results firmly establishes the suppression of parasitic
pockets as the primary mechanism for the strain-enhanced charge-to-spin
conversion in \ce{KV2Se2O}. The remaining discrepancy in the maximum CSE
($\sim 90\%$ vs.\ $\sim 96\%$) is attributed to the incomplete renormalization
of nearest-neighbor hopping anisotropies under the simplified uniform scaling. A
detailed discussion of the quantitative differences between the model and the DFT
results is provided in the Supplemental Material~\cite{supplement}.

\subsection{Tunneling Magnetoresistance Enhancement}

\begin{figure*}
  \centering
  \begin{overpic}[width=\textwidth]{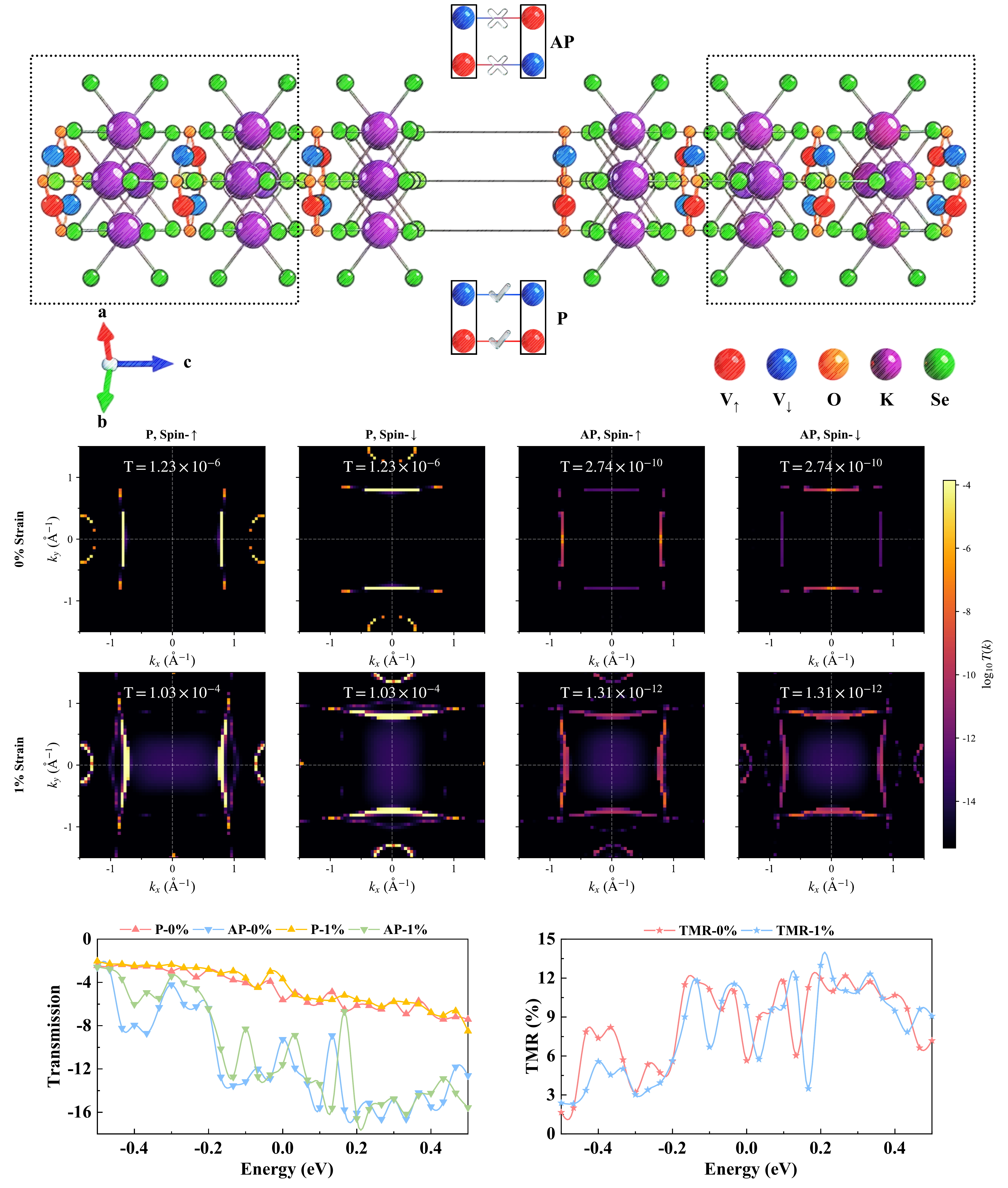}
    \put(2.5,98){\normalsize \textbf{(a)}}
    \put(2.5,63.5){\normalsize \textbf{(b)}}
    \put(42.5,63.5){\normalsize \textbf{(c)}}
    \put(2.5,44.5){\normalsize \textbf{(d)}}
    \put(42.5,44.5){\normalsize \textbf{(e)}}
    \put(2.5,22){\normalsize \textbf{(f)}}
    \put(42.5,22){\normalsize \textbf{(g)}}
  \end{overpic}
  \caption{(a) Schematic diagram of a dual-electrode MTJ structure based on \ce{KV2Se2O}.
    (b)--(c) Spin-$\overrightarrow{\bm{k}}_{\parallel}$-resolved transmission in the P and AP states of intrinsic \ce{KV2Se2O}.
    (d)--(e) The same as (b)--(c) but under 1\% strain, with spin-up (left) and spin-down (right) channels along the (001) crystallographic direction.
    (f) Energy-dependent transmission for the P and AP configurations.
    (g) Calculated TMR ratios as a function of energy.}
  \label{Fig5}
\end{figure*}

Beyond charge-to-spin conversion, the strain-induced elimination of parasitic Fermi
pockets also has profound consequences for tunneling magnetoresistance. To
illustrate this, we systematically investigate the modulation effect of strain on the
TMR in \ce{KV2Se2O}-based MTJ, using first-principles calculations combined with the
nonequilibrium Green's function (NEGF) method~\cite{NEGF1,NEGF2,NEGF3}. To rule out
the confounding effects of interface states and spin filtering introduced by an
intermediate barrier layer, we construct a MTJ structure with \ce{KV2Se2O} as both
electrodes and a vacuum layer serving as the tunneling barrier
[Figure~\hyperref[Fig5]{\ref{Fig5}(a)}], a modeling strategy widely adopted in
analogous studies~\cite{dong2022tunneling,obata2025pseudotunnel,yang2026altermagnetic}.
The configuration with parallel magnetization alignments of the two \ce{KV2Se2O}
electrodes is defined as the parallel (P) state, and that with antiparallel
magnetization alignments is defined as the antiparallel (AP) state.

We first comparatively analyze the momentum-space distributions of spin-up and
spin-down transport channels at the Fermi level in the MTJ with and without applied
strain [Figures~\hyperref[Fig5]{\ref{Fig5}(b)}--\hyperref[Fig5]{(e)}]. The calculation
results reveal that the parasitic Fermi pockets in \ce{KV2Se2O} are partially
suppressed under strain, accompanied by a pronounced enhancement in the P-state
tunneling conductance, which is consistent with the electronic structure conclusions
from the preceding analysis. Meanwhile, strain substantially reduces the overlap
between the two spin transport channels in momentum space, widening the gap in
tunneling probability between the P and AP configurations. According to the TMR
formula [Equation~\eqref{eq-tmr}], this enlarged discrepancy drives the TMR magnitude.

The pristine \ce{KV2Se2O}-based MTJ yields a TMR ratio of $10^{5}\%$, in good
agreement with results reported in existing literature~\cite{yang2026altermagnetic}.
Upon the application of strain modulation, the TMR magnitude rises to the order of
$10^{9}\%$, representing a remarkable enhancement of approximately four orders of
magnitude relative to the baseline value of the unstrained system. Finally, we
characterize the energy-dependent evolution of the MTJ transmission coefficient and
TMR near the Fermi level [Figures~\hyperref[Fig5]{\ref{Fig5}(f)}--\hyperref[Fig5]{(g)}].
The results demonstrate that a high TMR amplitude is preserved in the vicinity of the
Fermi level. This effect can effectively mitigate the detrimental impact of Fermi
level shifts induced by defects and doping in experiments on device performance,
thereby improving the process robustness of the device.

The concurrent enhancement of both CSE and TMR under strain originates from the same
microscopic mechanism: the removal of parasitic Fermi pockets that contribute unwanted
conduction channels. In the CSE context, these pockets dilute the spin-to-charge
conductivity ratio; in the MTJ context, they provide leakage channels in the AP state
that reduce the P/AP conductance contrast. This mechanism is governed by Fermi-surface
topology rather than precise on-site energetics. The NEGF transport calculations were
performed without a Hubbard $U$ correction, while the DFT calculations employed
$U_{\mathrm{eff}}=1.0$~eV~\cite{Hubbard_U2}; because the effective $U$ parameter
differs between projector augmented-wave (PAW)-based (VASP) and localized-basis-set (QuantumATK)
implementations~\cite{Hubbard_U_diff}, the strain values in the two frameworks are not
in strict one-to-one correspondence. The band-topology-driven
mechanism---suppression of parasitic pockets through lattice expansion---is
independent of the specific $U$ correction scheme.

\subsection{Materials Selectivity of the Pocket-Elimination Mechanism}

To assess whether the strain-driven pocket-elimination mechanism generalizes beyond
\ce{KV2Se2O}, we performed a comparative analysis of two isostructural analogues,
\ce{RbV2Te2O} and \ce{CsV2Te2O} (Supplemental Material~\cite{supplement}, Sec.~\ref{sec:pdos}).
These compounds share the same $P4/{\rm mmm}$ space group and $[C_2 \Vert C_{4z}]$
altermagnetic symmetry, but their electronic structures near $E_{\rm F}$ differ
markedly due to the larger spatial extent of Te $5p$ orbitals and the concomitant
increase in $p$--$d$ hybridization.

Table~\ref{tab:pdos_ratio} of the Supplemental Material~\cite{supplement} lists the orbital-resolved
PDOS ratio $|d_{xy}|/d_{xz}$ at $E_{\rm F}$ for all three compounds. In
\ce{KV2Se2O} (ratio $=0.51$), the $d_{xz}$ flat-band contribution dominates, and the
$|d_{xy}|$ pocket weight is selectively suppressed by strain. In \ce{RbV2Te2O}
(ratio $=1.24$), the pocket contribution already outweighs the flat-band
contribution, so strain simultaneously enhances both channels rather than selectively
removing the parasitic one. In \ce{CsV2Te2O} (ratio $=1.67$), the $d_{xz}$ flat-band
weight is nearly absent at $E_{\rm F}$, and the Fermi surface is dominated by broad,
multi-pocket features lacking well-defined orthogonal flat sheets.

These three contrasting cases establish a practical screening framework for
strain-engineered CSE in $d$-wave altermagnets. The ratio $|d_{xy}|/d_{xz}$ at
$E_{\rm F}$ serves as a predictive descriptor: effective strain enhancement requires
(i) both flat-band and pocket contributions to be present and distinguishable at
$E_{\rm F}$, and (ii) a strain response that preferentially suppresses the
pocket-generating hopping pathways. The fact that only \ce{KV2Se2O} satisfies both
criteria---while \ce{RbV2Te2O} and \ce{CsV2Te2O} fail for distinct, physically
transparent reasons---validates the selectivity of the mechanism and provides a
transferable design principle for screening future $d$-wave altermagnet candidates.

\setlength {\parskip} {0pt}
\section{Conclusions}\label{conc}
In summary, we have demonstrated that the CSE in the $d$-wave
altermagnet \ce{KV2Se2O} can be systematically enhanced to near the theoretical limit
of $100\%$ by in-plane equibiaxial tensile strain. Using effective model analysis and
DFT calculations, we established that residual elliptical Fermi pockets near $X$ and
$Y$---arising from band warping---act as parasitic channels that dilute the spin
conductivity. Strain engineering eliminates these pockets, restoring the ideal
orthogonal flat Fermi surfaces and boosting the CSE to a record $\sim 96\%$ at $4\%$
strain. While Lai \emph{et al.}~\cite{lai_d-wave_2025} achieved $\sim 98\%$ CSE through
electron doping, strain offers a more intrinsic route: it directly manipulates the
lattice to restore the flat-band geometry without introducing extrinsic carriers or
altering the Fermi level. Chemical doping, in contrast, inevitably introduces disorder
and shifts the system away from charge neutrality. Strain is a clean, continuously
tunable parameter, readily accessible in epitaxial thin-film platforms and compatible
with conventional semiconductor processing~\cite{ZHAO2026100441,AFM-RuO2-S}.
Furthermore, we uncovered an unconventional out-of-plane spin current component
$\sigma_{\Uparrow}^{\hat{\mathbf{n}}}$ that achieves a CSE of nearly $55\%$ at optimal
orientations, enabling field-free perpendicular magnetization switching.
Beyond charge-to-spin conversion, our magnetic-tunnel-junction calculations reveal
that the strain-driven elimination of parasitic Fermi pockets also yields a giant
enhancement of tunneling magnetoresistance. The TMR ratio rises from $10^{5}\%$ in the
unstrained junction to approximately $10^{9}\%$ under strain, representing a
four-order-of-magnitude increase and remaining robust against Fermi-level shifts.

The cross-material comparison presented above demonstrates that the
pocket-elimination mechanism is inherently materials-selective, governed by the
orbital-resolved PDOS ratio $|d_{xy}|/d_{xz}$ at $E_{\rm F}$. The screening criteria
identified here---validated against the full isostructural series \ce{KV2Se2O},
\ce{RbV2Te2O}, and \ce{CsV2Te2O}---provide a transferable design principle for
identifying $d$-wave altermagnet candidates amenable to strain-enhanced CSE beyond
the present materials series.

This strain-enabled dual functionality---near-unity CSE and giant TMR---highlights
the central role of the flat-band geometry and establishes a general
materials-by-design strategy for maximizing $\mathcal{T}$-odd spin currents and
magnetotransport in $d$-wave altermagnets.
While demonstrated here on \ce{KV2Se2O}, this approach is applicable to a
broad class of altermagnets where similar pocket structures limit the charge-to-spin
conversion, paving the way toward next-generation high-efficiency spintronic devices.

\section{Calculation Methods}\label{Calc}
DFT calculations were performed with the
Vienna \textit{ab initio} Simulation Package (VASP)~\cite{25}. The PAW pseudopotentials~\cite{26} were employed together with the
Perdew-Burke-Ernzerhof (PBE) generalized gradient approximation for the
exchange-correlation functional~\cite{27}. A plane-wave kinetic energy cutoff of
$520$~eV was used. All structures were fully relaxed until residual forces fell
below $0.01$~eV/\AA\ and the energy convergence reached $10^{-7}$~eV. For Brillouin
zone integration during relaxation, a $\Gamma$-centered $9 \times 9\times 5$
Monkhorst-Pack $k$-mesh was applied; for self-consistent electronic structure
calculations the mesh was densified to $13\times13\times9$. To account for the
correlated nature of the V $3d$ electrons, an effective Hubbard $U_{\mathrm{eff}} =
1.0$~eV was introduced, consistent with previous spectroscopic studies of this
system~\cite{jiang_metallic_2025,Hubbard_U1,Hubbard_U2}.

In-plane equibiaxial tensile strain was applied by scaling the in-plane lattice
constants $a$ and $b$ by a factor $(1+\varepsilon)$, where $\varepsilon$ ranges from
0\% to 5\%. For each strain value, the in-plane lattice parameters were fixed while
the out-of-plane lattice constant $c$ and all internal atomic positions were fully
relaxed. This procedure mimics the experimental configuration of coherent epitaxial
growth on a (001)-oriented substrate.

Maximally localized Wannier functions (MLWFs) were constructed from the V $3d$,
O $2p$, Se $4p$, and K $3d_{x^2-y^2}$ orbitals using the Wannier90
code~\cite{marzari_maximally_2012}, faithfully reproducing the DFT band structure
near the Fermi level. The Wannier interpolation was performed on a dense $k$-point
mesh of $200\times200\times200$ to ensure convergence of the transport integrals. The
electric-field-induced spin currents were evaluated within the linear-response
framework employing the Wannier-Linear-Response
package~\cite{gonzalez-hernandez_efficient_2021,29-SST2,30-WLR1}.

Within the constant relaxation-time approximation, the zero-temperature
$\mathcal{T}$-odd spin conductivity tensor is given by the Kubo
formula~\cite{31-PhysRevB.77.165117,32-T-odd} (see Supplemental
Material~\cite{supplement} for the explicit expression). The charge-to-spin
conversion efficiency (CSE) is defined as $\mathrm{CSE} =
|\sigma_{ij}^{k}/\sigma_{xx}|$, reflecting the ratio of the generated spin current
to the applied charge current. A constant broadening parameter $\eta =
\SI{10}{\milli\electronvolt}$ was used throughout the transport calculations.

To elucidate the microscopic mechanism underlying the strain-enhanced CSE, we
constructed a two-orbital tight-binding model in the basis of the V $d_{xz}/d_{yz}$
(flat band) and $d_{xy}$ (elliptical pocket) orbitals. The spin-up and spin-down
sectors are decoupled in the absence of SOC and are related by the $[C_2 \Vert
C_{4z}]$ symmetry. The full set of hopping parameters, obtained by fitting to the DFT
band structure, is provided in the Supplemental Material~\cite{supplement}. To
simulate the effect of tensile strain, the NNN hopping amplitudes responsible for the
elliptical pockets are uniformly scaled by a factor $\alpha$ ranging from 1.00
(pristine) to 0.80.

At the device level, a two-terminal structure with \ce{KV2Se2O}/vacuum layer/\ce{KV2Se2O} sandwiched
between two semi-infinite \ce{KV2Se2O} leads is employed to investigate the TMR effect.
The transport performance of the MTJ is calculated using the DFT with NEGF formalism~\cite{NEGF4} as implemented in QuantumATK~\cite{QATK1}.
The Brillouin zone is sampled using Monkhorst-Pack $k$-point meshes of $11 \times 11 \times 100$ for the \ce{KV2Se2O}/vacuum layer/\ce{KV2Se2O}
device along the [001] direction. The TMR of the MTJ is given by:
\begin{equation}
  {\rm TMR}=\frac{T_{\rm P}(\overrightarrow{\bm{k}_{\parallel}})-T_{\rm AP}(\overrightarrow{\bm{k}_{\parallel}})}{T_{\rm AP}(\overrightarrow{\bm{k}_{\parallel}})} \label{eq-tmr}
\end{equation}
where $\overrightarrow{\bm{k}_{\parallel}}$ is the wave vector transverse to the transport direction, and $T_{\rm P}$ and $T_{\rm AP}$ represent
the transmission coefficients of the MTJ in P and AP states, respectively.

\section*{Acknowledgments}
The authors would like to acknowledge the financial support from National Key Research and Development Program of China (Grant No. 2022YFA1602701), and National Natural Science Foundation of China (Grants No. 12574131, No. 12327806, and No. 12227806).

\section*{Conflict of Interest}
The authors declare no conflict of interest.

\section*{Data Availability Statement}
The data that support the findings of this study are available from the corresponding author upon reasonable request.

\singlespacing

%

\end{document}